\documentclass[%
 reprint,
superscriptaddress,
 amsmath,amssymb,
 prb,
]{revtex4-2}

\usepackage{graphicx}
\usepackage{dcolumn}
\usepackage{bm}
\allowdisplaybreaks
\usepackage{amsfonts}
\usepackage{amsmath}
\usepackage{txfonts}
\usepackage{amssymb}
\usepackage{amsbsy} 
\usepackage{graphicx}
\usepackage{dcolumn}
\usepackage{bm}
\allowdisplaybreaks
\usepackage[breaklinks]{hyperref}
\hypersetup{colorlinks=true, linkcolor=blue, citecolor=blue, filecolor=blue, urlcolor=blue}
\graphicspath{{figures}}
\usepackage{dsfont}
\usepackage{tikz}
\usetikzlibrary{quantikz}

\begin{document}


\title{Direct Probe of Topology and Geometry of Quantum States on IBM Q}

\author{Tianqi Chen}
\email{tqchen@nus.edu.sg}
\affiliation{Department of Physics, National University of Singapore, Singapore 117551}

\author{Hai-Tao Ding}
\email{htding@smail.nju.edu.cn}
\affiliation{Department of Physics, National University of Singapore, Singapore 117551}
\affiliation{National Laboratory of Solid State Microstructures and School of Physics, Nanjing University, Nanjing 210093, China}
\affiliation{Collaborative Innovation Center of Advanced Microstructures, Nanjing 210093, China}

\author{Ruizhe Shen}
\affiliation{Department of Physics, National University of Singapore, Singapore 117551}

\author{Shi-Liang Zhu}
\affiliation {Key Laboratory of Atomic and Subatomic Structure and Quantum Control (Ministry of Education), Guangdong Basic Research Center of Excellence for Structure and Fundamental Interactions of Matter, and School of Physics, South China Normal University, Guangzhou 510006, China}
\author{Jiangbin Gong}
\email{phygj@nus.edu.sg}
\affiliation{Department of Physics, National University of Singapore, Singapore 117551}
\affiliation{Centre for Quantum Technologies, National University of Singapore, Singapore 117543}
\affiliation{Joint School of National University of Singapore and Tianjin University, International Campus of Tianjin University, Binhai New City, Fuzhou 350207, China}

\begin{abstract}
The concepts of topology and geometry are of critical importance in exploring exotic phases of quantum matter.  Though they have been investigated on various experimental platforms, to date a direct probe of topological and geometric properties on a universal quantum computer even for a minimum model is still in vain.  In this work, we first show that a density matrix form of the quantum geometric tensor (QGT) can be explicitly re-constructed from Pauli operator measurements on a quantum circuit.  We then propose two algorithms, suitable for IBM quantum computers,  to directly probe QGT.  The first algorithm is a variational quantum algorithm particularly suitable for Noisy Intermediate-Scale Quantum (NISQ)-era devices, whereas the second one is a pure quantum algorithm based on quantum imaginary time evolution.  Explicit results obtained from IBM Q simulating a Chern insulator model are presented and analysed. Our results indicate that transmon qubit-based universal quantum computers have the potential to directly simulate and investigate topological and geometric properties of a quantum system.

\end{abstract}

\maketitle

\section{Introduction}
The geometry and topology are fundamental concepts in many branches of the modern physics~\cite{nakahara2018geometry}, ranging from condensed matter physics to astrophysics. The interplay between these concepts was demonstrated in
the two-dimensional (2D) anomalous quantum Hall insulator, where the topological gapless edge state is characterized by the Chern number~\cite{haldane1988model,thouless1982quantized}. This topological effect relies on the non-trivial geometry of the energy bands, and the geometric properties of them are fully encoded by the quantum geometric tensor~\cite{provost1980riemannian,kolodrubetz2017geometry,ma2010abelian} (QGT). The imaginary part of QGT is the Berry curvature, which is responsible for various topics, such as anomalous Hall transport~\cite{sundaram1999wave,nagaosa2010anomalous}, Aharonov-Bohm effect~\cite{aharonov1959significance}, and topological quantum matters~\cite{hasan2010colloquium,qi2011topological,zhang2018topological}, whereas the real part of QGT corresponds to the quantum metric, which reflects the distance between two nearby quantum states in Hilbert space and bears numerous fascinating physical phenomena, {including quantum phase transition~\cite{zanardi2007information,ren2024identifying}, semi-classical dynamics~\cite{bleu2018effective}, orbital magnetism~\cite{Gao2015Geometrical,Piechon2016Geometric}, topological quantum phases~\cite{Roy2014Band,Lim2015geometry,Palumbo2018revealing,ding2024non,ding2020tensor}, etc. Furthermore, QGT is closely related to superfluidity in flat bands, which is proportional to superfluid weight in Hermitian~\cite{julku2016geometric} and non-Hermitian cases~\cite{he2021geometry}.

Experimental studies of QGT have been reported in a number of artificially engineered quantum systems}~\cite{liao2021experimental,tan2021experimental,chen2022synthetic,asteria2019measuring,yu2022quantum,zheng2022measuring,lysne2023quantum}, including the superconducting qubits~\cite{TanYu2019} and qutrits~\cite{Tan2018topological}, the nitrogen-vacancy center in diamond~\cite{Yu2020experimental,Yu2022experimental}, exciton-polaritons in the planar micro-cavity~\cite{Gianfrate2020measurement}, and cold atoms in optical lattices\cite{YiPan2023Extracting}.
In addition, most recently, there are additional theoretical proposals of observing QGT in other mesoscopic systems such as plasmonic lattice~\cite{Cuerda2023observation,Cuerda2023pseudospinorbit} and photonic systems\cite{Bleu2018measuring,Hu2023generalized}. Nevertheless, these experimental platforms rely on indirect measurements such as performing periodic time evolution~\cite{TanYu2019},  leading to extended processing time and erroneous outcomes. Anticipating the power and promises of universal quantum computers in the quantum simulation of topological quantum matter,  developing a direct and robust means to probe QGT, suitable for the setting of universal quantum computers, will be a necessary step. Indeed, with technological advances,  there have been excellent progresses on the use of superconducting qubits-based quantum computers for studying exotic phases in condensed matter physics~\cite{smith2019simulating,RahmaniZhang2020,google2020hartree, mi2022time,Pouyan2022,mi2022time,koh2022simulation,FreyRachel2022,kim2023evidence,koh2023observation,Chen2023robust,Shen2023observation,ma2023limitations,xiang2023simulating,Xiang2024long,mei2020digital,Shtanko2023tjn,GoogleQuantumAI2024Dynamics}. In this work, we present a direct scheme to probe all the components of the QGT on an IBM Q quantum processor, using the seminal Qi-Wu-Zhang (QWZ) model~\cite{qi2006topological}.  Specifically, based on our proposed density matrix formalism of the QGT of the ground state, one may perform a direct measurement of Pauli operators on the quantum state and then reconstruct the full QGT for a wide range of parameters in the momentum space. To implement this direct route,  we introduce two distinct approaches: one being an entirely quantum algorithm by employing imaginary time evolution to obtain the ground state without any classical pre-processing, and the other one being a variational optimization algorithm performed on a parameterized quantum circuit (PQC). Both approaches are shown to be feasible in the measurements of the QGT on IBM Q,  executed parallely on  quantum circuits for a chosen range of system parameters. This feature is markedly different from Ref.~\cite{TanYu2019} where only a single qubit is executed at a time for each set of parameters. Our algorithms are hence more efficient and feasible on a universal quantum computer, without requiring any \textit{a priori} information of the ground state to calibrate the initial state~\cite{TanYu2019} before the simulation.  Interestingly,  the quantum imaginary time evolution algorithm is less robust to the current-stage noise and gate errors.

\section{Methods: Calculation of the Abelian QGT from a quantum computer}
\label{sec:methods}

As the results obtained from any IBM quantum processor are counts of the computational basis of qubits, the explicit state vector could not be extracted from the data, and only the density matrix can be reconstructed via state tomography~\cite{Cramer2010efficient}. To this end, we derived the equation to extract both the Abelian and non-Abelian quantum geometric tensor.

Here, we show how to explicitly obtain Abelian quantum geometric tensor in the projection operator formalism. We consider a general $2\times2$ Hamiltonian $H(k_{\mu},k_{\nu})$, with the ground state denoted as $|\psi_g\rangle$, and the excited state denoted as $|\psi_e\rangle$. The corresponding projection operators are $P_g=|\psi_g\rangle \langle \psi_g|$, and $P_e=|\psi_e\rangle \langle \psi_e|$. Then, it is found that
\begin{equation}
\label{eq:lhs}
\frac{\partial P_g}{\partial_{k_\mu}}\cdot P_{e} \cdot \frac{\partial P_g}{\partial_{k_\nu}}=\left| \psi_g\right\rangle\left\langle\partial_\mu \psi_g\right| \psi_e\left\rangle\right\langle \psi_e\left|\partial_{\nu}\psi_g\right\rangle\left\langle \psi_g\right|.
\end{equation}
The Abelian quantum geometric tensor $Q_{\mu\nu}$ for the ground state is
\begin{equation}
\label{eq:rhs}
\begin{aligned}
Q_{\mu \nu}&=\left\langle\partial_\mu \psi_g\right|\left(I_2-|\psi_g\rangle\langle \psi_g|\right)\left| \partial_\nu \psi_g \right\rangle  \\
&=\left\langle\partial_\mu \psi_g | \psi_e\right\rangle\left\langle \psi_e \mid \partial_\nu \psi_g\right\rangle.
\end{aligned}
\end{equation}
where $I_2$ is $2\times 2$ identity matrix.
Now, with the above two relations from Eq.~\eqref{eq:lhs} and Eq.~\eqref{eq:rhs}, we can then derive the quantum geometric tensor from projection operator
\begin{equation}
    \frac{\partial P_g}{\partial_{k_\mu}}\cdot P_{e} \cdot \frac{\partial P_g}{\partial_{k_\nu}}=Q_{\mu\nu}P_g.
\end{equation}
The real part of the quantum geometric tensor $Q_{\mu \nu}$ defines the quantum metric tensor, and its imaginary part is the Berry curvature \cite{Berry1984quantal,WilczekShapere1989,XiaoNiu2010}
\begin{equation}
    Q_{\mu\nu}=g_{\mu\nu}-\frac{i}{2}F_{\mu\nu}.
\end{equation}
Then we are able to calculate both the quantum metric tensor and the Berry curvature from a pair of equations of the projection operators as Eq.~\eqref{eq:metricandcurvature}.

Because the matrix representation of projector operators ($P_g$ and $P_e$) from Eq.~\eqref{eq:metricandcurvature} is of size $2 \times 2$, i.e. each matrix consists of four elements, both the values of $g_{\mu \nu}$ or $F_{\mu \nu}$ are determined by four equations, and therefore we compute the values of each $g_{\mu \nu}$ and $F_{\mu \nu}$ by averaging over all values obtained from all four equations.

\section{Results}
\subsection{Calculation of the QGT from Pauli observables}
\begin{figure*}[t]
	\centering
	\includegraphics[width=2.0\columnwidth,draft=false]{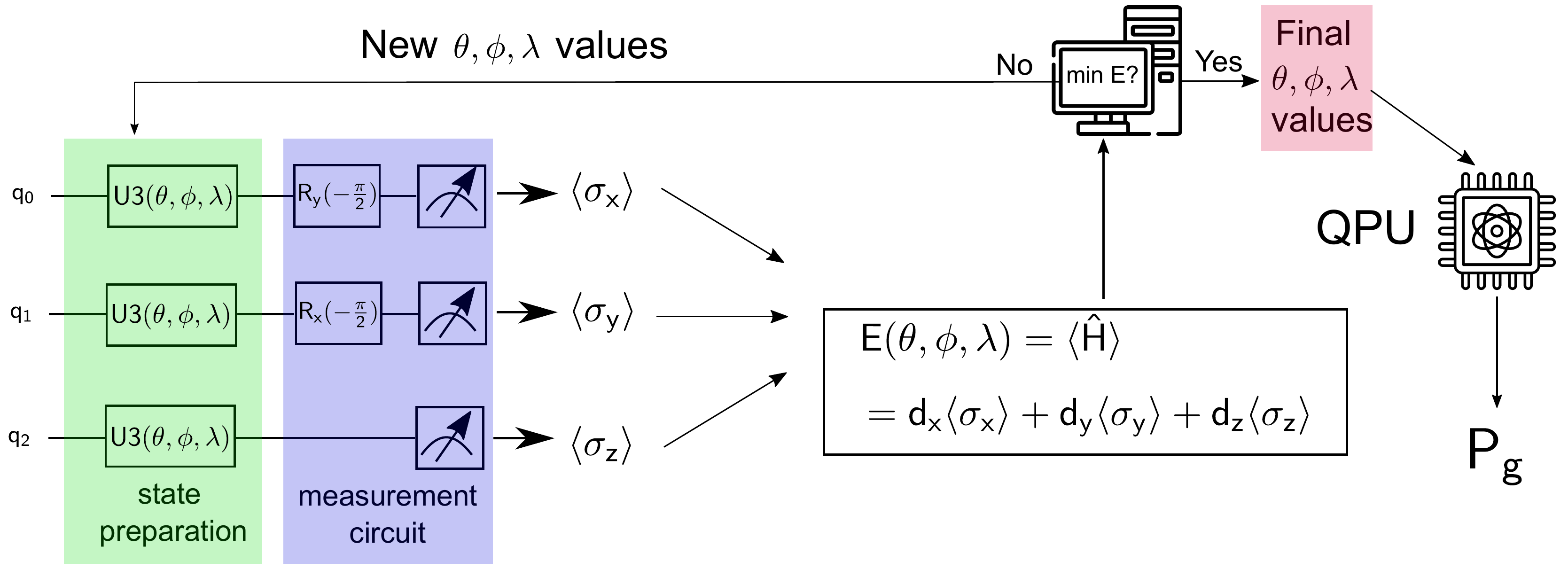}
	\caption{{Illustration of variational quantum algorithm (VQA) for the preparation of the projector representation of the ground state from the Hamiltonian in Eq.~\eqref{eq:cherninsulatorhamiltonian}: $P_g=| \psi_g\rangle\langle \psi_g|$.} To obtain the target ground state $|\psi_g\rangle$, one employs a parametrized quantum circuit (PQC) minimizes the total energy $\langle \hat{\mathcal{H}} \rangle$ with VQA, which is composed of the state preparation part (green) as well as the measurement part (blue). After the optimization, the resulting parameters ($\theta,\phi,\lambda$) for $U3$ gates are obtained, and the final PQCs are then executed on real machines. We remark that the convention for the $U3$ gates ($U(\theta,\phi,\lambda)$) follows the one from Qiskit SDK~\cite{Qiskit}. }
	\label{fig:VQEillustration}
\end{figure*}

\subsubsection{Chern insulator and quantum geometric tensor (QGT)}
Here, we provide a brief introduction of a general two-band Hamiltonian, as well as the expression of quantum geometric tensor. For simplicity, we consider the Qi-Wu-Zhang model~\cite{qi2006topological}, of which the Hamiltonian can be explicitly expressed as
\begin{align}
\label{eq:cherninsulatorhamiltonian}
    &\hat{\mathcal{H}}=d_x\sigma_x+d_y\sigma_y+d_z\sigma_z,
\end{align}
where the Bloch vector $\textbf{d}=(d_x,d_y,d_z)=(\sin k_x,\sin k_y,m-\cos k_x-\cos k_y)$, the $2$D momentum space parameters are denoted as $\mathbf{k}=(k_x,k_y)$ ($k_x,k_y \in [0, 2\pi]$), and $m$ is the tunable Zeeman strength. $\sigma_i$ ($i=x,y,z$) are the Pauli matrices. For this Hamiltonian, when $m \in (0,2)$, it is a $2$D Chern insulator, and trivial band insulator when $m>2$. The topological invariant to characterize this topological phase transition is the first Chern number~\cite{thouless1982quantized}. We focus on the ground state of the Hamiltonian from Eq.~\eqref{eq:cherninsulatorhamiltonian}, which we denote it as $|\psi_{g}\rangle$. The quantum geometric tensor can be defined as
\begin{align}
\label{eq:QGTgeneralexpression}
Q_{\mu \nu}&=\langle\partial_{\mu} \psi_{g}| \partial_{\nu}\psi_{g}\rangle-\langle\partial_{\mu} \psi_{g}|\psi_{g}\rangle\langle\psi_{g}|\partial_\nu\psi_{g}\rangle   \\
&=g_{\mu\nu}-\frac{i}{2}F_{\mu\nu}.
\end{align}
with $\{\mu,\nu\}=\{k_x,k_y\}$.
The real and also symmetric part of the QGT defines the quantum metric
\begin{equation}
\begin{aligned}
    g_{\mu\nu}=&\frac{1}{2}(\langle\partial_\mu \psi_g \mid \partial_\nu \psi_g\rangle+\langle\partial_\nu \psi_g \mid \partial_\mu \psi_g\rangle  \\
    &-\langle\partial_\mu \psi_g \mid \psi_g\rangle \langle \psi_g \mid\partial_\nu \psi_g\rangle-\langle\partial_\nu \psi_g \mid \psi_g\rangle \langle \psi_g \mid\partial_\mu \psi_g\rangle).
\end{aligned}
\end{equation}
which essentially defines the distance~\cite{provost1980riemannian} between two neighboring states 
$|\psi(\mathbf{k})\rangle$ and $|\psi(\mathbf{k}+d\mathbf{k})\rangle$ [$d\mathbf{k}=(dk_x, dk_y)$] in Hilbert space.

The imaginary and also anti-symmetric part of it corresponds to the Berry curvature
\begin{equation}
F_{\mu\nu}=i(\langle\partial_\mu \psi_g \mid \partial_\nu \psi_g\rangle-\langle\partial_\nu \psi_g \mid \partial_\mu \psi_g\rangle),
\end{equation}
which defines the geometric phase difference of a wave function undergoing parallel transport along a closed path in the parameter space.

From the QGT, one can calculate the first Chern number from the integration of the Berry curvature over the first Brillouin zone
\begin{align}
    \label{eq:chernnumberexpression}
    &\mathcal{C}=\frac{1}{2\pi}\int_{\mathbb{T}^2}F_{xy}\,dk_x dk_y,
\end{align}
which characterizes the topological phase transition of the Hamiltonian in Eq.~\eqref{eq:cherninsulatorhamiltonian}.

\subsubsection{Expression of the ground state density matrix in terms of Pauli observables}
\label{sec:QGTPauli}
To obtain the QGT, the explicit state vector form for the ground state ($|\psi_g\rangle$) is needed for each set of parameters [Eq.~\eqref{eq:QGTgeneralexpression}].  However, for any quantum algorithm, the final state obtained has a global phase which in principle cannot be measured.  This global phase issue indicates that there is no need to reconstruct the whole statevector with the global phase. That is, there must be a way to re-express the QGT without the global phases of the quantum states parameterized by the system parameters.  This is indeed achieved here.  Inspired by the idea of quantum state tomography~\cite{Choo2018Measurement}, we first perform Pauli operator $\langle \sigma_x \rangle$, $\langle \sigma_y \rangle$, and $\langle \sigma_z \rangle$ measurements to extract the density matrix of the ground state $|\psi_g\rangle$, where the global phase is canceled and the density matrix element is determined. For any density matrix of a single-qubit ground state $P_g=|\psi_g \rangle \langle \psi_g |$ expressed in the matrix form:
\begin{align}
\label{eq:densitymatrixrho}
    &P_g=\left[\begin{array}{cc}
    \alpha & \beta \\
    \gamma & \delta    
    \end{array}\right],
\end{align}
the elements could all be explicitly determined by Pauli measurements of a same quantum state $|\psi_g\rangle$. The diagonal components can be reconstructed by 
\begin{align}
    \alpha &= \langle \uparrow | \psi_g \rangle \langle \psi_g | \uparrow \rangle =|\langle \uparrow |\psi_g\rangle |^2,\\ \nonumber
    \delta &= \langle \downarrow | \psi_g \rangle \langle \psi_g | \downarrow \rangle =|\langle \downarrow |\psi_g\rangle |^2,
\end{align}
while for the off-diagonal components of $P_g$, they are 
\begin{align}
    &\beta= \frac{1}{2}\langle \sigma_x \rangle - \frac{i}{2}\langle \sigma_y \rangle, \\ \nonumber 
    &\gamma=\frac{1}{2}\langle \sigma_x \rangle + \frac{i}{2}\langle \sigma_y \rangle.
\end{align}
More importantly, all  the components of the metric tensor $g_{\mu \nu}$ and the Berry curvature $F_{\mu\nu}$ can then be obtained via solving the following equations
\begin{align}
\label{eq:metricandcurvature}
&\frac{1}{2}\left(\frac{\partial P_g}{\partial_{\mu}}\cdot P_{e} \cdot \frac{\partial P_g}{\partial_{\nu}}+\frac{\partial P_g}{\partial_{\nu}}\cdot P_{e} \cdot \frac{\partial P_g}{\partial_{\mu}}\right)=g_{\mu\nu}P_g,    \\ \nonumber
&i\left(\frac{\partial P_g}{\partial_{\mu}}\cdot P_{e} \cdot \frac{\partial P_g}{\partial_{\nu}}-\frac{\partial P_g}{\partial_{\nu}}\cdot P_{e} \cdot \frac{\partial P_g}{\partial_{\mu}}\right)=F_{\mu\nu}P_g,
\end{align}
{where $P_e=I - P_g$ is the density matrix for the excited state, $\partial_{\mu/\nu}\equiv\partial_{k_\mu/k_{\nu}}$, and $\mu, \nu=\{x,y\}$. Also, we have denoted $|\partial_{\mu/\nu}\psi_g\rangle=\frac{\partial}{\partial_{\mu/\nu}}|\psi_{g}\rangle$ and $\langle\partial_{\mu/\nu}\psi_g|=\frac{\partial}{\partial_{\mu/\nu}}\langle\psi_{g}|$.} A more detailed derivation of how to obtain the above equations in the realm of density matrix is provided in Methods. In addition, this approach can also be extended to the non-Abelian case where the ground states have degeneracies (see Appendix).

\subsection{Quantum algorithm}
\label{sec:VQAChern}
\subsubsection{Variational quantum algorithm for ground state preparation and QGT}
\begin{figure}[t]
	\centering
			\begin{quantikz}
                    &\lstick{$\ket{ \uparrow}_{\sf A_0}$}&\gate[wires=2]{\tilde{\sf U}} \gategroup[6,steps=2,style={dashed,
					rounded corners,fill=blue!20, inner xsep=2pt},
					background]{{\sf state preparation}} &\gate[]{\sf Post-selection} &\qw &\qw\\
				&\lstick{$\ket{\uparrow}_{\rm P_0}$}&\qw&\qw&\gate{\sf R_y(-\pi/2)} \gategroup[1,steps=2,style={dashed,rounded
				corners,fill=red!20, inner
				xsep=2pt},background,label style={label
				position=below,anchor=north,yshift=-0.2cm}]{{\sf measurement of $\sf \langle\sigma_x\rangle$}}  &\meter{}\\
                    &\lstick{$\ket{\uparrow}_{\sf A_1}$}&\gate[wires=2]{\tilde{\sf U}}&\gate[]{\sf Post-selection}&\qw&\qw\\
				&\lstick{$\ket{\uparrow}_{\sf P_1}$}&\qw&\qw&\gate{\sf R_x(\pi/2)} \gategroup[1,steps=2,style={dashed,rounded
				corners,fill=red!20, inner
				xsep=2pt},background,label style={label
				position=below,anchor=north,yshift=-0.2cm}]{{\sf measurement of $\sf \langle\sigma_y\rangle$}} &\meter{}\\
				&\lstick{$\ket{\uparrow}_{\sf A_2}$}&\gate[wires=2]{\tilde{\sf U}}&\gate[]{\sf Post-selection}&\qw&\qw\\
				&\lstick{$\ket{\uparrow}_{\sf P_2}$}&\qw&\qw&\qw \gategroup[1,steps=2,style={dashed,rounded
				corners,fill=red!20, inner
				xsep=2pt},background,label style={label
				position=below,anchor=north,yshift=-0.2cm}]{{\sf measurement of $\sf \langle\sigma_z\rangle$}} &\meter{}
			\end{quantikz}
	\caption{{Illustration of the quantum circuits for quantum imaginary time evolution (ITE) and obtaining projection operators of the ground state $P_g$.} The circuit consists of the part for the state preparation with a unitary operator $U$ [see Eq.~\eqref{u}] and the post-selection (blue), followed by the measurement part to extract the full density matrix $P_g$. The initial state is $\ket{\uparrow\uparrow}=|\uparrow\rangle_{\rm A} \otimes |\uparrow \rangle_{\rm P}$. A total of $6$ qubits are used where $(A_0,P_0)$ are for measuring $\langle \sigma_x \rangle$, $(A_1,P_1)$ for $\langle \sigma_y \rangle$, and $(A_2,P_2)$ for $\langle \sigma_z \rangle$. For each pair of the qubits, the first qubit (labelled with $|\uparrow\rangle_{\rm A_{i}} (i=0,1,2)$) is the ancilla qubit, which requires the final post-selection of $\ket{\uparrow}$, and the second one (labelled with $|\uparrow\rangle_{\rm P_i}(i=0,1,2)$) is the physical qubit. After the post-selection, the result from the physical qubit is normalized as $e^{-tH_{\rm TB}}\ket{\uparrow}/\left\| e^{-tH_{\rm TB}}\ket{\uparrow}\right\| $. The explicit gate decomposition of $U$ is shown in Fig.~\ref{fig:unitary operator circuit} in Appendix.}
	\label{fig:ITEcircuit}
\end{figure}
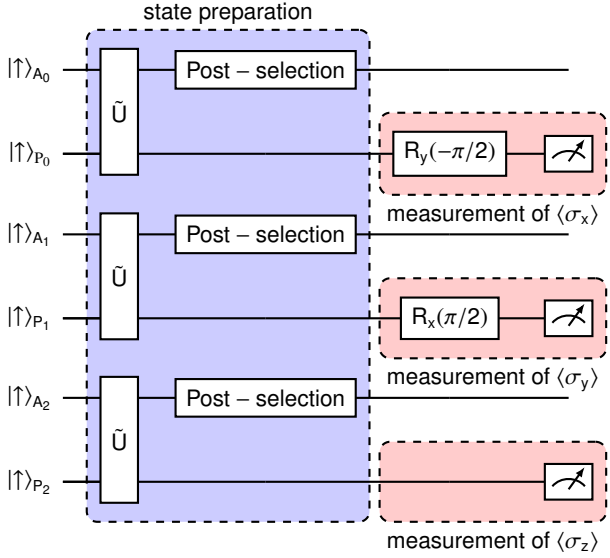
Here, we introduce a scheme based on the variational quantum algorithm~\cite{Yuan2019theoryofvariational} for the ground state preparation.  This scheme is most feasible for current NISQ-era devices due to various types of noise.  With this algorithm we prepare the ground state for the two-band model and subsequently obtain the quantum geometric tensor as well as its Chern number on IBM Q via direct measurement of different Pauli operator observables.  Full descriptions of post-measurement processing or feed-forward procedures on a classical computer to determine the outcomes are described in Methods.

In Fig.~\ref{fig:VQEillustration}, we illustrate the main idea of preparing the ground state of the Hamiltonian from Eq.~\eqref{eq:cherninsulatorhamiltonian}.  For each set of parameter $(k_x,k_y)$, a set of three qubits are used and initialized in spin-up's. We use a single-layered $U3=U3(\theta,\phi,\lambda)$ gates~\cite{Qiskit} where $\theta,\phi,\lambda$ are gate parameters to be optimized (see Appendix for the detailed definition of $U3$ gate) to form a parameterized quantum circuit (PQC) for the state preparation (green), followed by the measurement circuit (blue) for all $x$, $y$ and $z$ Pauli operators. The total Hamiltonian energy $\langle \hat{\mathcal{H}} \rangle$ is then optimized to obtain a PQC which faithfully represents the ground state $| \psi_g \rangle$. Finally, the resulting PQC is sent to the IBM Q for execution, and the ground state density matrix $P_g=|\psi_g\rangle \langle \psi_g|$ can be obtained by solving a simple pair of equations (see Methods).

\begin{figure*}[t]
	\centering
	\includegraphics[width=2.0\columnwidth,draft=false]{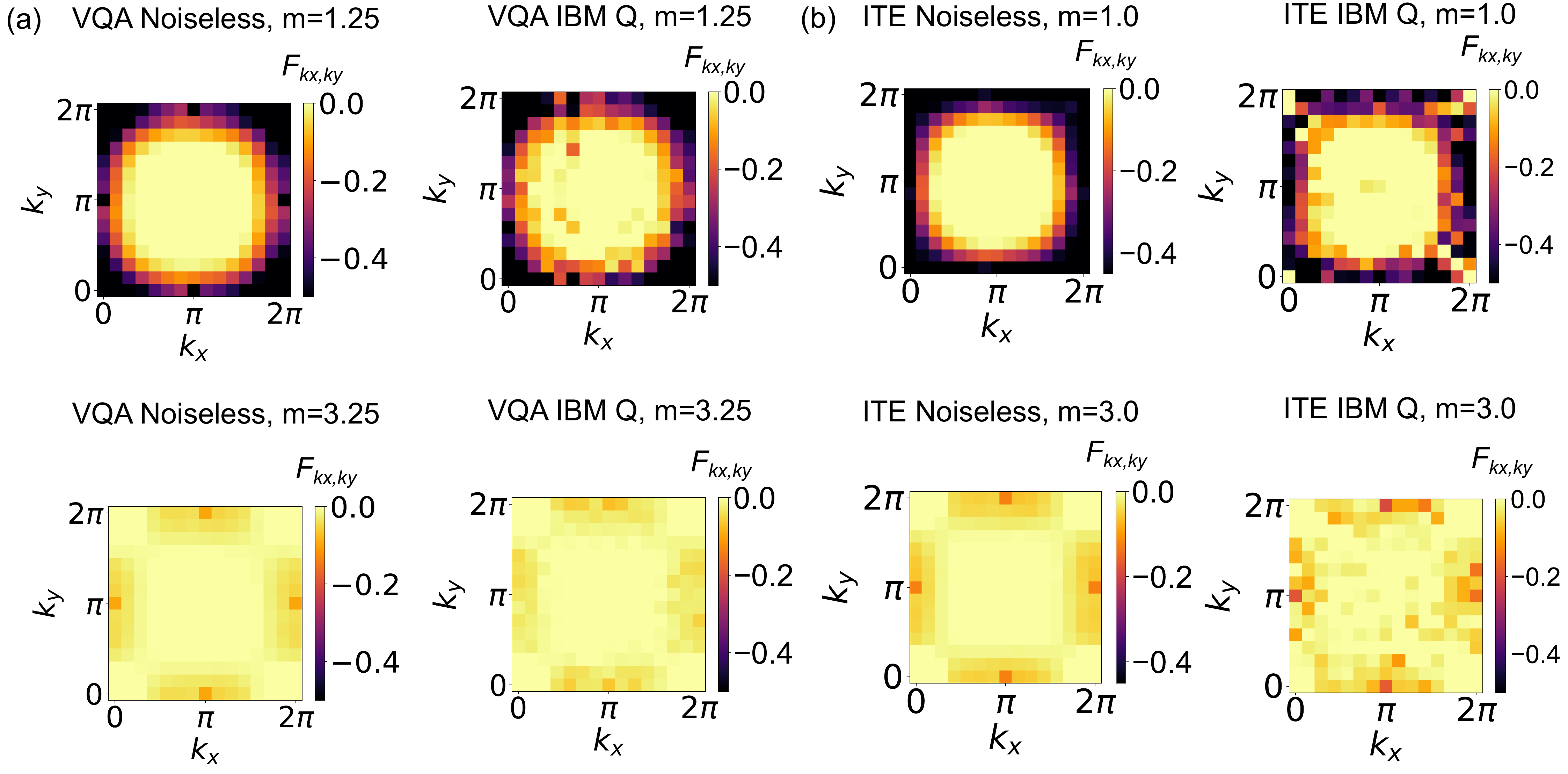}
	\caption{ {Comparison of the calculation of Berry curvature $F_{k_x,k_y}$ using VQA and quantum ITE:} (a) $F_{k_x,k_y}$ calculated using VQA and (b) $F_{k_x,k_y}$ calculated using quantum ITE. The noisy results were obtained on IBM Q device \textit{ibmq\_algiers}. For all panels, we have chosen both cases from the non-trivial topological phase ($m<2$) and the trivial phase ($m>2$), and we have set the increment for both momenta $k_x,k_y$ as $\delta=0.04\pi$. }
	\label{fig:VQEBerryCurvatureresults}
\end{figure*}

\subsubsection{Quantum ITE algorithm for ground state preparation and QGT}
\label{sec:QGTResults}
We further explore the possibility of probing QGT without resorting to any optimization procedure beforehand. To that end, we perform the quantum imaginary time evolution (ITE)~\cite{Mcardle2019variational,Motta2020Determining,Nishi2021implementation,Sun2021Quantum,Kamakari2022Digital,Mao2023Measurement,Chen2023efficient} with post-selection on an additional ancilla qubit to obtain the ground state of the Hamiltonian on IBM Q.  In contrast with our variational quantum algorithm in the previous section, our approach is based on a pure quantum algorithm to obtain the ground state of an arbitrary Hamiltonian using the quantum ITE.

In principle, our aim is to physically perform the following operation on a quantum computer: 
\begin{align}
\label{eq:ITEforGS}
&\lim_{\tau \rightarrow \infty} e^{-\tau H}|\psi_0\rangle/|| e^{-\tau H}|\psi_0\rangle || \rightarrow |\psi_{g}\rangle,
\end{align}
where $|\psi_0\rangle$ is an arbitrary initial state, and $|\psi_{g} \rangle$ is the non-degenerate ground state of the Hamiltonian $\hat{\mathcal{H}}$. Here, we propose to simulate the imaginary time evolution by the two-band Hamiltonian $\hat{\mathcal{H}}$ as $U_{\rm TB}=e^{-\tau \hat{\mathcal{H}}_{\rm TB}}$ (Here, $\hat{\mathcal{H}}_{\rm TB}=\hat{\mathcal{H}}$ from Eq.~\eqref{eq:cherninsulatorhamiltonian} ) when $\tau \rightarrow \infty$, which can be transformed to an enlarged unitary operator $\tilde{U}$~\cite{LinPollmannReal2021}. First, we embed $U_{\rm TB}$ into a $4\times 4$ matrix $\mathcal{M}$ as follows:

\begin{align}
\label{u}
	&\mathcal{M}=\left[\begin{array}{cc}
		uU_{\rm TB} & B \\
		C & D
	\end{array}\right].
\end{align}
where $u^{-2}$ is equal to the maximum eigenvalue of $U_{\rm TB}^{\dagger}U_{\rm TB}$, and $C=\sqrt{I-u^{2}U_{\rm TB}^{\dagger}U_{\rm TB}}$. $B$ and $D$ can be arbitrary~\cite{LinPollmannReal2021}. Then, $\tilde{U}$ can be obtained by the QR decomposition~\cite{Terashima2005nonunitary,LinPollmannReal2021,Chen2022high,Shen2023observation}:
\begin{align}
\label{supussh}
	&\mathcal{M}=\tilde{U}R=\left[\begin{array}{cc}
		uU_{\rm TB} & I\\    
		C& I
	\end{array}\right],
\end{align}
where we replace $B$ and $D$ with the identity matrix $I$, and $R$ is an upper triangular matrix. We choose $B,D$ to be the identity matrix $I$ for simplicity~\cite{Chen2022high,Shen2023observation}. As a result, the QR decomposition does not change the upper-left block, and therefore the operator $\tilde{U}$ is the one to be implemented on a quantum circuit. Since the Hilbert space is now enlarged, an additional ancilla qubit is introduced and therefore the initial state becomes
\begin{align}
    &|\psi\rangle = |\psi_0 \rangle \otimes |\uparrow \rangle_A
\end{align}
where $\psi_0$ is the physical state, and the ancilla qubit is initialized in the spin-up state. 

The circuit to realize this algorithm is depicted in Fig.~\ref{fig:ITEcircuit}. In contrast to the previous VQA approach, for each measurement of the Pauli operator $x,y,z$, there is a post-selection of spin-up on each ancilla qubit:
\begin{align}
    \label{eq:postselection}
    &\langle \uparrow |_A \tilde{U} \left( |\psi_0 \rangle \otimes |\uparrow \rangle_A \right)=U_{\rm TB} | \psi_o \rangle
\end{align}
such that the imaginary-time evolution unitary operator $U_{\rm TB}$ is acted on the physical initial state, i.e.,  after its action, the final outcome state is the target ground state.

\subsection{Results from noisy simulations on IBM Q}
\begin{figure}[t]
	\centering
	\includegraphics[width=1.0\columnwidth,draft=false]{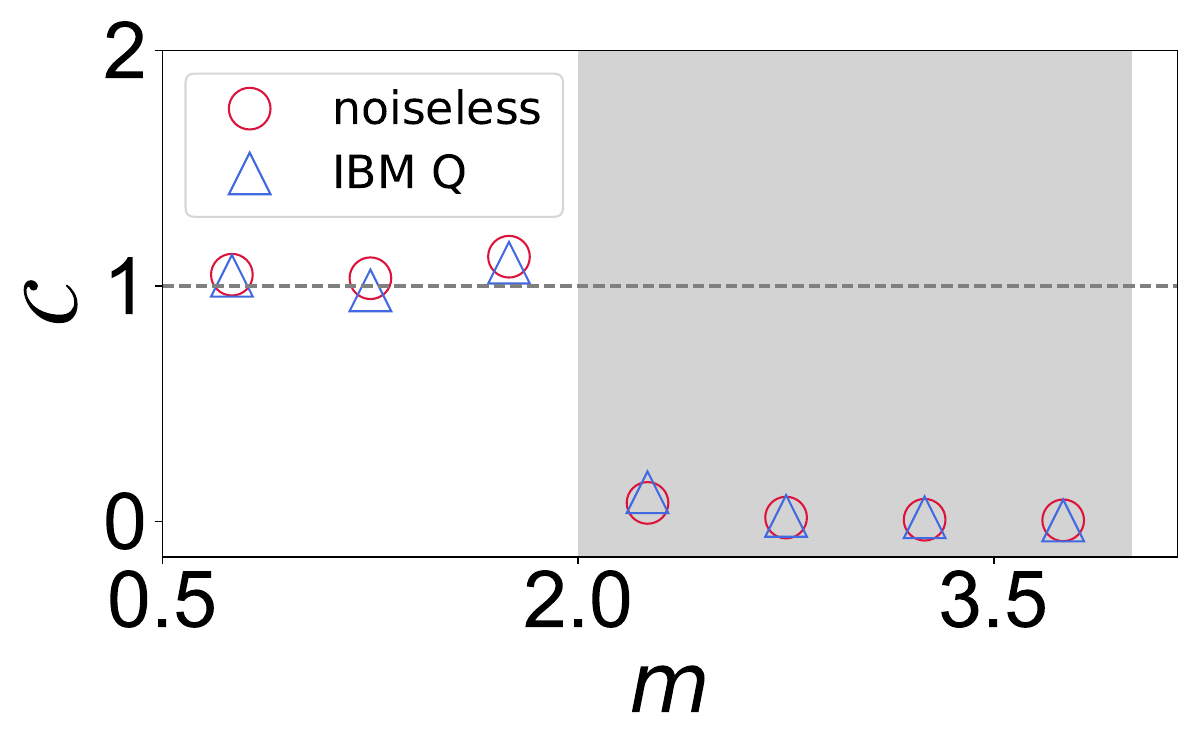}
	\caption{{Comparison of the calculation of Chern number with using VQA:} The data in red (blue) color corresponds to the noiseless (real IBM device \texttt{ibm\_algiers} backend) results. The shaded grey area indicates the trivial phase. }
	\label{fig:VQAChernresults}
\end{figure}

We show in Fig.~\ref{fig:VQEBerryCurvatureresults} the results of the Berry curvature $F_{k_x, k_y}$ of the two-band model for both algorithms of VQA ($m=1.25$ in the non-trivial topological phase, and $m=3.25$ in the trivial phase ) and ITE ($m=1.0$ in the non-trivial topological phase, and $m=3.0$ in the trivial phase).     Details of our calculations of the Berry curvature $F_{k_x, k_y}$ can be found in Methods.  For VQA, it is found that the real device results [Fig.~\ref{fig:VQEBerryCurvatureresults}(a)] are consistent with those obtained from noiseless classical simulation of the quantum circuits for both phases. For all cases, shallow circuits with only one layer of $U3$ gates are utilized [Fig.~\ref{fig:VQEillustration}]. From the Berry curvature results, one obtains the Chern number $\mathcal{C}$ as a function of $m$ in Fig.~\ref{fig:VQAChernresults} by numerically performing the integration from Eq.~\eqref{eq:chernnumberexpression} over a $15 \times 15$ grid in the first Brillouin zone. Similar to the Berry curvature results, both real IBM Q device results and noiseless simulations are consistent with each other, and a sharp transition between the non-trivial topological phase and the trivial phase is observed around $m=2$. Our findings clearly show that for a NISQ-era device, a robust and high-fidelity QGT can be obtained via variational quantum algorithms. Because the state preparation as well as the measurement circuit is shallow, our approach could further support downstream operations on QGT itself on a quantum computer, plus other most recent feed-forward approaches such as mid-circuit measurements~\cite{Smith2023deterministic,Koh2023measurement} and dynamic circuits~\cite{Corcoles2021exploiting} on a transmon qubit-based quantum computer.

\begin{figure}[h]
	\centering
	\includegraphics[width=1.0\columnwidth,draft=false]{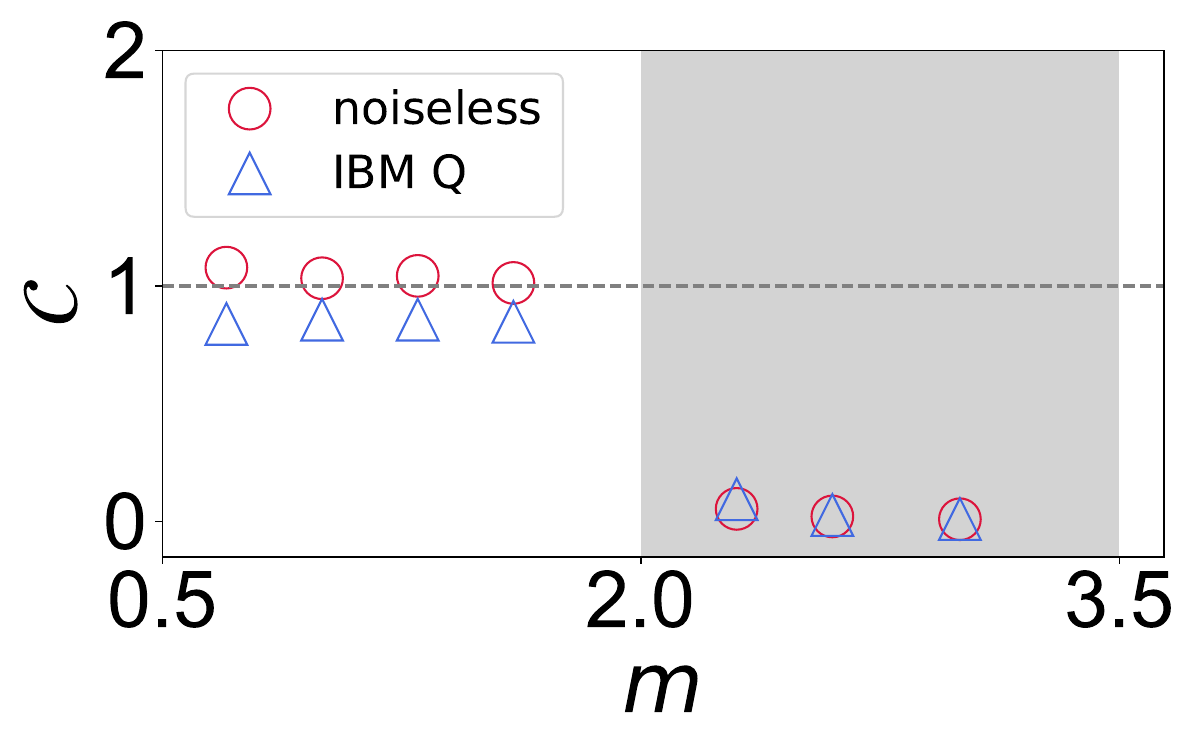}
	\caption{{Comparison of the calculation of Chern number using ITE:} The data in red (blue) corresponds to the noiseless \texttt{Aer} simulator (real backend \texttt{ibm\_algiers}) results. The shaded grey area indicates the trivial phase. }
	\label{fig:ITEChernresults}
\end{figure}

On the other hand, for the results obtained using ITE [Fig.~\ref{fig:VQEBerryCurvatureresults}(b)],  the Berry curvature obtained from the real IBM Q device has some deviations for both topological phase ($m=1.0$) and the trivial phase ($m=3.0$) at certain parameter points.   The real device performance for the Berry curvature is hence not as high quality as compared that obtained with VQA [Fig.~\ref{fig:VQEBerryCurvatureresults}(a)].  This is partially because the VQA algorithm does not involve CNOT gates from the decomposition of the unitary operator [see Appendix]. 
 Nevertheless, the results for the Chern number on IBM Q via ITE are qualitatively consistent with the noiseless results [Fig.~\ref{fig:ITEChernresults}]. As a consequence,  the topological phase transition could be still clearly captured even in the noisy simulations on IBM Q.   In particular,  the numerical integration of the Berry curvature on a real device is robust for the topological phase with $m<2$ [Fig.~\ref{fig:ITEChernresults}] even in the presence of some clearly inaccurate results in the Berry curvature. This is possible because inaccuracies  in the Berry curvature results can cancel with each other when being integrated.  For the trivial phase, we note that both the real IBM Q device results and the noiseless results are more consistent with each other than the topologically nontrivial phase.  Indeed, in these topologically trivial cases the intermediate level of noise and the numerical integration of the Berry curvature itself both help to yield an almost exact zero value for the Chern number.

To summarize our observations, the ITE approach presented here does yield the qualitatively correct simulation of QGT on a universal quantum computer.  As a pure quantum algorithm, the ITE approach has the potential to show quantum advantage when studying different phenomena in condensed matter physics.  This is especially so when the error correction on a quantum computer is available and the device noise is further brought down. In addition to the above Berry curvature and Chern number simulations from two different approaches, we have shown our results for the quantum metric tensor ($g_{k_x,k_x}, g_{k_y,k_y}, g_{k_x,k_y}$) using both VQA and ITE approaches in the Appendix as well.

{To conclude this section, we remark that from a technical perspective, for both algorithms of VQA and ITE, the quantum circuits were executed simultaneously on the quantum computer for a variety of varying parameters, in contrast with previous methods~\cite{TanYu2019} where the procedure was conducted sequentially. For details of the simulations on IBM Q, see Appendix.}

\section{Discussion}
In this work, we have introduced a direct method to probe all the elements of quantum geometric tensor (QGT), using the two-band model in the momentum space as a working example.  Utilizing a prevalent NISQ-era IBM Q quantum processor, we propose two distinct algorithms: firstly, an entirely quantum algorithm employing imaginary time evolution to prepare the ground state, and secondly, a variational optimization algorithm operating on a parameterized quantum circuit (PQC) for ground state preparation. We have conducted a comparative analysis of their efficacy, finding that current NISQ-era device through the variational optimization algorithm can better capture QGT signatures across a topological phase transition.  The quantum imaginary time evolution algorithm yields less accurate results due to prevailing noise and gate errors, with room for enhancement in the near future when quantum computer noise levels are further suppressed with more logical qubits or via error correction \cite{KimEvidence2023,EveredHighfidelity2023,BluvsteinLogical2023}. Notably, both algorithms adopted here
are executed on quantum circuits for a variety of parameters in the momentum space at the same time, eliminating the need for a priori Hamiltonian ground state information for calibration. Building on the success of the current study, it is of considerable interest to simulate  the timely object of QGT in non-Hermitian systems~\cite{Bergholtz2021exceptional,Lin2023topological,Okuma2023nonhermitian,ShenLee2023YL} as well as in open quantum systems~\cite{Breuer2007theory,Michishita2022dissipation,ChenPoletti2020steady,ChenPoletti2021thermodynamic}, which are still at their infancy in current literature.  The theory part of this work has also laid down a solid foundation towards  the simulation of non-Abelian QGT~\cite{ma2010abelian,ding2022extracting} on quantum computers.

\section{Acknowledgement}
T.~C. and H.-T.~D. contributed equally to this project. This work is supported by the Singapore
National Research Foundation via the Project No.~NRF2021-
QEP2-02-P09, as well as the support by the National
Research Foundation, Singapore and A*STAR under its
CQT Bridging Grant. Hai-Tao Ding and Shi-Liang Zhu are supported by the National Natural Science Foundation of China under the Grant No. 12074180 and the National Key Research and Development Program of China under Grant No. 2022YFA1405300. We acknowledge the use of IBM Quantum services for this work via Singapore’s Quantum Engineering Programme (QEP). The views expressed are those of the authors, and do not reflect the official policy or position of IBM or the IBM Quantum team. Part of the classical simulations of quantum circuits from this article was partially performed on resources of the National Supercomputing Centre, Singapore (\url{https://www.nscc.sg/}), and on the National University of Singapore (NUS)'s high-performance computing facilities. The data and the code which supports the findings of this study are available from the corresponding author upon request.

\bibliography{ref}

\newpage
\onecolumngrid
\appendix

\section{Calculation of the non-Abelian QGT in the density matrix formalism}
This method can be extended to the Hamiltonian with the degenerate subspace, the geometric quantity to characterize it is non-Abelian quantum geometric tensor. We now show how to explicitly obtain the non-Abelian QGT in the projection operator formalism. We consider a $4\times 4$ Hamiltonian $H(k_{\mu},k_{\nu})$ parameterized by arbitrary parameters $(k_{\mu}, k_{\nu})$, degenerate ground states are $|\psi_1\rangle$ and $|\psi_2\rangle$, and excited states are $|\psi_3\rangle$ and $|\psi_4\rangle$ respectively. In this case, quantum geometric tensor $Q$ is a $4\times 4$ matrix
\begin{equation}
Q =\left(\begin{array}{cc}
Q_{\mu\mu} & Q_{\mu\nu} 
\vspace{1ex}\\
Q_{\nu\mu} & Q_{\nu\nu}
\end{array}\right) =\left(\begin{array}{cccc}
Q_{\mu\mu}^{11} & Q_{\mu\mu}^{12} & Q_{\mu\nu}^{11} & Q_{\mu\nu}^{12}
\vspace{1ex}\\
Q_{\mu\mu}^{21} & Q_{\mu\mu}^{22} & Q_{\mu\nu}^{21} & Q_{\mu\nu}^{22} 
\vspace{1ex}\\
Q_{\nu\mu}^{11} & Q_{\nu\mu}^{12} & Q_{\nu\nu}^{11} & Q_{\nu\nu}^{12}
\vspace{1ex}\\
Q_{\nu\mu}^{21} & Q_{\nu\mu}^{22} & Q_{\nu\nu}^{21} & Q_{\nu\nu}^{22}
\end{array}\right).
\end{equation}
with
\begin{equation}
Q_{\mu \nu}^{ij}=\langle\partial_{\mu} \psi_{i}|\left(1-|\psi_1\rangle\langle \psi_1|-|\psi_2\rangle\langle \psi_2|\right)| \partial_{\nu}\psi_{j}\rangle.
\end{equation}
The relation between the non-Abelian quantum geometric tensor, the non-Abelian quantum metric, and non-Abelian Berry curvature is 
\begin{equation}
Q_{\mu\nu}=g_{\mu\nu}-\frac{i}{2}F_{\mu\nu}.
\end{equation}
Now $g_{\mu\nu}$ and $F_{\mu\nu}$ are all $2\times2$ matrices.
\begin{equation}
\begin{aligned}
\label{eq:g and F}
g_{\mu\nu}&=(Q_{\mu\nu}+Q_{\mu\nu}^{\dagger})/2, \\
F_{\mu\nu}&=i(Q_{\mu\nu}-Q_{\mu\nu}^{\dagger}).
\end{aligned}
\end{equation}
The total $g$ and $F$ are $4\times4$ matrices
\begin{equation}
g=\left(\begin{array}{cc}
g_{\mu\mu} & g_{\mu\nu} 
\vspace{1ex}\\
g_{\nu\mu} & g_{\nu\nu}
\end{array}\right) =\left(\begin{array}{cccc}
g_{\mu\mu}^{11} & g_{\mu\mu}^{12} & g_{\mu\nu}^{11} & g_{\mu\nu}^{12}
\vspace{1ex}\\
g_{\mu\mu}^{21} & g_{\mu\mu}^{22} & g_{\mu\nu}^{21} & g_{\mu\nu}^{22} 
\vspace{1ex}\\
g_{\nu\mu}^{11} & g_{\nu\mu}^{12} & g_{\nu\nu}^{11} & g_{\nu\nu}^{12}
\vspace{1ex}\\
g_{\nu\mu}^{21} & g_{\nu\mu}^{22} & g_{\nu\nu}^{21} & g_{\nu\nu}^{22}
\end{array}\right).
\end{equation}
\begin{equation}
F=\left(\begin{array}{cc}
F_{\mu\mu} & F_{\mu\nu} 
\vspace{1ex}\\
F_{\nu\mu} & F_{\nu\nu}
\end{array}\right) =\left(\begin{array}{cccc}
F_{\mu\mu}^{11} & F_{\mu\mu}^{12} & F_{\mu\nu}^{11} & F_{\mu\nu}^{12}
\vspace{1ex}\\
F_{\mu\mu}^{21} & F_{\mu\mu}^{22} & F_{\mu\nu}^{21} & F_{\mu\nu}^{22} 
\vspace{1ex}\\
F_{\nu\mu}^{11} & F_{\nu\mu}^{12} & F_{\nu\nu}^{11} & F_{\nu\nu}^{12}
\vspace{1ex}\\
F_{\nu\mu}^{21} & F_{\nu\mu}^{22} & F_{\nu\nu}^{21} & F_{\nu\nu}^{22}
\end{array}\right).
\end{equation}
We define the following projection operators as
\begin{equation}
    \begin{aligned}
        P_1&=|\psi_1\rangle \langle \psi_1|,  \\
        P_2&=|\psi_2\rangle \langle \psi_2|,  \\
        P_g&=|\psi_1\rangle \langle \psi_1| + |\psi_2\rangle \langle \psi_2|,  \\
        P_e&=|\psi_3\rangle \langle \psi_3| + |\psi_4\rangle \langle \psi_4|.
    \end{aligned}
\end{equation}
Then it is found that 
\begin{equation}
\begin{aligned}
    \frac{\partial P_g}{\partial_{k_\mu}}\cdot P_{e} \cdot \frac{\partial P_g}{\partial_{k_\nu}}&=Q_{\mu\nu}^{11}|\psi_1\rangle \langle \psi_1|+Q_{\mu\nu}^{12}|\psi_1\rangle \langle \psi_2|+Q_{\mu\nu}^{21}|\psi_2\rangle \langle \psi_1|+Q_{\mu\nu}^{22}|\psi_2\rangle \langle \psi_2|.
\end{aligned}
\end{equation}
For the non-Abelian quantum metric
\begin{equation}
\begin{aligned}
&\frac{1}{2}(\frac{\partial P_g}{\partial_{k_\mu}}\cdot P_{e} \cdot \frac{\partial P_g}{\partial_{k_\nu}}+\frac{\partial P_g}{\partial_{k_\nu}}\cdot P_{e} \cdot \frac{\partial P_g}{\partial_{k_\mu}})=g_{\mu\nu}^{11}|\psi_1\rangle \langle \psi_1|+g_{\mu\nu}^{12}|\psi_1\rangle \langle \psi_2|+g_{\mu\nu}^{21}|\psi_2\rangle \langle \psi_1|+g_{\mu\nu}^{22}|\psi_2\rangle \langle \psi_2|.
\end{aligned}
\end{equation}
For the non-Abelian Berry curvature
\begin{equation}
\begin{aligned}
&i(\frac{\partial P_g}{\partial_{k_\mu}}\cdot P_{e} \cdot \frac{\partial P_g}{\partial_{k_\nu}}-\frac{\partial P_g}{\partial_{k_\nu}}\cdot P_{e} \cdot \frac{\partial P_g}{\partial_{k_\mu}})=F_{\mu\nu}^{11}|\psi_1\rangle \langle \psi_1|+F_{\mu\nu}^{12}|\psi_1\rangle \langle \psi_2|+F_{\mu\nu}^{21}|\psi_2\rangle \langle \psi_1|+F_{\mu\nu}^{22}|\psi_2\rangle \langle \psi_2|.
\end{aligned}
\end{equation}
We can use the projection operator to cancel irrelevant terms, to get $g_{\mu\nu}^{11}$, $g_{\mu\nu}^{22}$, $F_{\mu\nu}^{11}$, $F_{\mu\nu}^{22}$
\begin{equation}
\label{eq:Abelianprojectionoperator}
\begin{aligned}
P_1\left\{\frac{1}{2}(\frac{\partial P_g}{\partial_{k_\mu}}\cdot P_{e} \cdot \frac{\partial P_g}{\partial_{k_\nu}}+\frac{\partial P_g}{\partial_{k_\nu}}\cdot P_{e} \cdot \frac{\partial P_g}{\partial_{k_\mu}})\right\}P_1&=g_{\mu\nu}^{11}P_1,  \\
P_2\left\{\frac{1}{2}(\frac{\partial P_g}{\partial_{k_\mu}}\cdot P_{e} \cdot \frac{\partial P_g}{\partial_{k_\nu}}+\frac{\partial P_g}{\partial_{k_\nu}}\cdot P_{e} \cdot \frac{\partial P_g}{\partial_{k_\mu}})\right\}P_2&=g_{\mu\nu}^{22}P_2,  \\
P_1\left\{i\frac{\partial P_g}{\partial_{k_\mu}}\cdot P_{e} \cdot \frac{\partial P_g}{\partial_{k_\nu}}-\frac{\partial P_g}{\partial_{k_\nu}}\cdot P_{e} \cdot \frac{\partial P_g}{\partial_{k_\mu}}\right\}P_1&=F_{\mu\nu}^{11}P_1,  \\
P_2\left\{i\frac{\partial P_g}{\partial_{k_\mu}}\cdot P_{e} \cdot \frac{\partial P_g}{\partial_{k_\nu}}-\frac{\partial P_g}{\partial_{k_\nu}}\cdot P_{e} \cdot \frac{\partial P_g}{\partial_{k_\mu}}\right\}P_2&=F_{\mu\nu}^{22}P_2.
\end{aligned}
\end{equation}
But one cannot get $g_{\mu\nu}^{12}$, $g_{\mu\nu}^{21}$, $ F_{\mu\nu}^{12}$ and $F_{\mu\nu}^{12}$ from the above method, that is to say,
\begin{equation}
P_1\left\{\frac{1}{2}(\frac{\partial P_g}{\partial_{k_\mu}}\cdot P_{e} \cdot \frac{\partial P_g}{\partial_{k_\nu}}+\frac{\partial P_g}{\partial_{k_\nu}}\cdot P_{e} \cdot \frac{\partial P_g}{\partial_{k_\mu}})\right\}P_2=g_{\mu\nu}^{12}|\psi_1\rangle \langle \psi_2|.
\end{equation}
Then we define two ground states as
\begin{equation}
\begin{aligned}
|\psi_M\rangle&=\frac{1}{\sqrt{2}}(|\psi_1\rangle+|\psi_2\rangle),  \\
|\psi_N\rangle&=\frac{1}{\sqrt{2}}(|\psi_1\rangle+i|\psi_2\rangle).
\end{aligned}
\end{equation}
Corresponding projection operators $P_M=|\psi_M\rangle \langle \psi_M|$, $P_N=|\psi_N\rangle \langle \psi_N|$. Eq.~\eqref{eq:Abelianprojectionoperator} is derived based on subspace $\left\{|\psi_1\rangle, |\psi_2\rangle\right\}$. If we consider new subspace $\left\{|\psi_M\rangle, |\psi_N\rangle\right\}$, Eq.~\eqref{eq:Abelianprojectionoperator} will be changed to
\begin{equation}
\begin{aligned}
P_M\left\{\frac{1}{2}(\frac{\partial P_M}{\partial_{k_\mu}}\cdot P_{e} \cdot \frac{\partial P_M}{\partial_{k_\nu}}+\frac{\partial P_M}{\partial_{k_\nu}}\cdot P_{e} \cdot \frac{\partial P_M}{\partial_{k_\mu}})\right\}P_M&=g_{\mu\nu}^{MM}P_M, \\
P_N\left\{\frac{1}{2}(\frac{\partial P_N}{\partial_{k_\mu}}\cdot P_{e} \cdot \frac{\partial P_N}{\partial_{k_\nu}}+\frac{\partial P_N}{\partial_{k_\nu}}\cdot P_{e} \cdot \frac{\partial P_N}{\partial_{k_\mu}})\right\}P_N&=g_{\mu\nu}^{NN}P_N,  \\
P_M\left\{i\frac{\partial P_M}{\partial_{k_\mu}}\cdot P_{e} \cdot \frac{\partial P_M}{\partial_{k_\nu}}-\frac{\partial P_M}{\partial_{k_\nu}}\cdot P_{e} \cdot \frac{\partial P_M}{\partial_{k_\mu}}\right\}P_M&=F_{\mu\nu}^{MM}P_M, \\
P_N\left\{i\frac{\partial P_N}{\partial_{k_\mu}}\cdot P_{e} \cdot \frac{\partial P_N}{\partial_{k_\nu}}-\frac{\partial P_N}{\partial_{k_\nu}}\cdot P_{e} \cdot \frac{\partial P_N}{\partial_{k_\mu}}\right\}P_N&=F_{\mu\nu}^{NN}P_N.  \\
\end{aligned}
\end{equation}
Then we can get $g_{\mu\nu}^{MM}$, $g_{\mu\nu}^{NN}$, $F_{\mu\nu}^{MM}$ and $F_{\mu\nu}^{NN}$. Actually, $g_{\mu\nu}^{ij}$ and $F_{\mu\nu}^{ij}$ can be expressed as
\begin{equation}
\label{eq:nonAbelianqgt}
\begin{aligned}
g_{\mu \mu}^{i j}&=\frac{2 i g_{\mu \mu}^{MM}+2 g_{\mu \mu}^{NN}-(1+i)\left(g_{\mu \mu}^{i i}+g_{\mu \mu}^{j j}\right)}{2 i},   \\
g_{\mu \nu}^{i j}&=\frac{2 i g_{\mu \nu}^{MM}+2 g_{\mu \nu}^{NN}-(1+i)\left(g_{\mu \nu}^{i i}+g_{\mu \nu}^{j j}\right)}{2 i},   \\
F_{\mu \mu}^{ij}&=\frac{2 i F_{\mu \mu}^{MM}+2 F_{\mu \mu}^{NN}-(1+i)\left(F_{\mu \mu}^{ii}+F_{\mu \mu}^{jj}\right)}{2 i},  \\
F_{\mu \nu}^{ij}&=\frac{2 i F_{\mu \nu}^{MM}+2 F_{\mu \nu}^{NN}-(1+i)\left(F_{\mu \nu}^{ii}+F_{\mu \nu}^{jj}\right)}{2 i}.
\end{aligned}
\end{equation}
with $i,j=\left\{1,2\right\}$. From Eq.~\eqref{eq:nonAbelianqgt}, we can get all $g_{\mu\nu}^{ij}$ and $F_{\mu\nu}^{ij}$.
In a word, all the components of non-Abelian quantum metric and Berry curvature can be extracted from the projection operators.

\section{Additional results for computing quantum metric tensor $g_{\mu \nu}$ on IBM Q using variational quantum algorithm}

\begin{figure*}[h]
	\centering
	\includegraphics[width=1.0\columnwidth,draft=false]{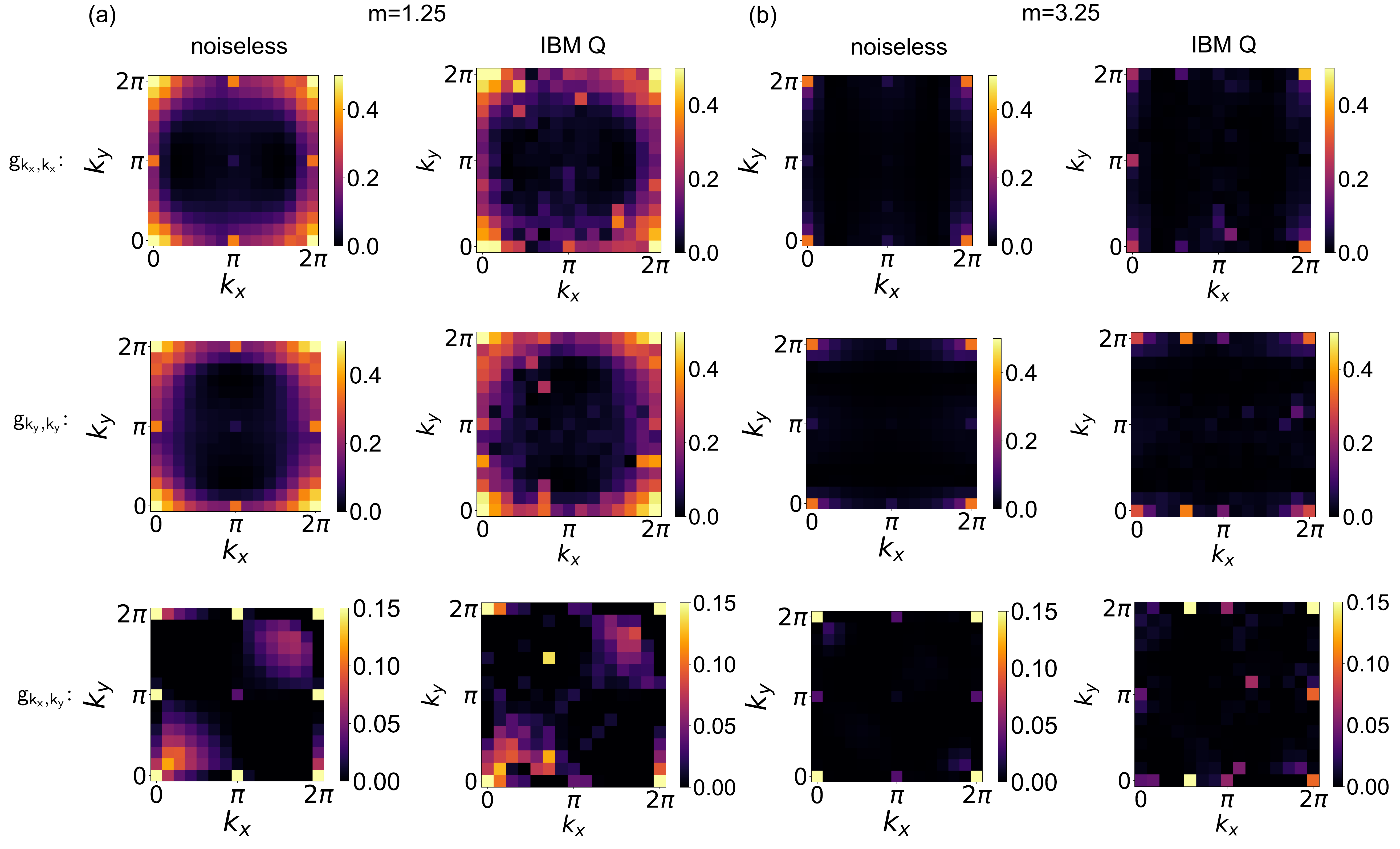}
	\caption{VQA results of quantum metric tensor $g_{k_x,k_y}$, $g_{k_y,k_y}$, and $g_{k_x,k_y}$ for $m=1.25$ and $m=3.25$. Both noiseless simulations and real IBM Q device simulations are presented here.}
	\label{fig:QGT_VQE}
\end{figure*}

In Fig.~\ref{fig:QGT_VQE}, using the variational quantum algorithm (VQA) approach, we show the results of quantum metric tensor $g_{\mu,\nu}$ ($\mu,\nu=k_x,k_y$) as obtained from Eq.~\eqref{eq:g and F}, where $k_x,k_y \in [0, 2\pi]$. It is found that the results obtained from noisy simulations on IBM Q are qualitatively consistent with those from the noiseless Aer simulations~\cite{Qiskit} for both topological and trivial phases. Specifically, we notice that even for the noiseless results, there are a few anomalous points (in orange color) centered around $0, \pi$ and $2\pi$ for the topological phase, and $0,2\pi$ for the trivial phase. This is intrinsically due to the numerically error in the calculation: around these parameters in the first Brillouin zone, certain ground state projector $P_g$ elements undergo a sharp transition and are quite close to zero, and since we calculate the metric tensor values by dividing over those particular matrix elements from $P_g$, it gives such an abnormal points in the results. We remark that the results may be improved in the future when more logical qubits are available, and the gate errors are further suppressed.

\section{Details of digital simulations on the IBM Q device}

\begin{figure*}[h]
 \centering
 \includegraphics[width=1.0\columnwidth,draft=false]{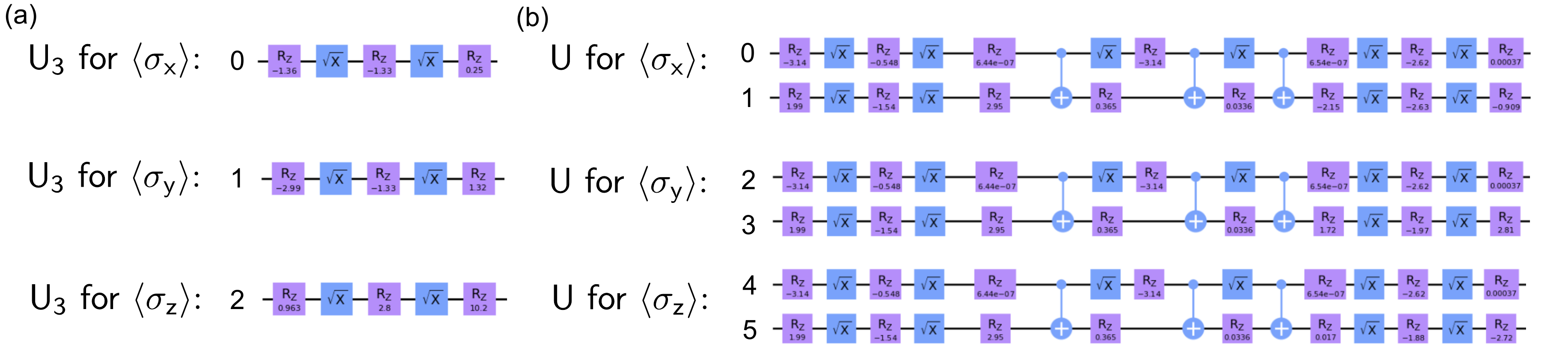}
	\caption{Unitary operators transpiled into basic gates on IBM Q for: (a) variational quantum algorithm. The $U3$ gates from Fig.~\ref{fig:VQEillustration} for the measurement of $\langle \sigma_x \rangle$, $\langle \sigma_y \rangle$, and $\langle \sigma_z \rangle$. (b) The $U$ operators from Fig.~\ref{fig:ITEcircuit}. For all Pauli operator measurements here, the circuit consists of at most $3$ CNOT gates, i.e., an almost constant-depth quantum circuit. For both panels, we only show the transpiled circuits for obtaining the ground state projection operator $P_g=| \psi_g \rangle \langle \psi_g |$ at $k_x=0,k_y=0$ for simplicity, as for parameters considered in this work, the circuit depths are mostly similarly.  }
	\label{fig:unitary operator circuit}
\end{figure*}

\begin{figure*}[h]
	\centering
	\includegraphics[width=1.0\columnwidth,draft=false]{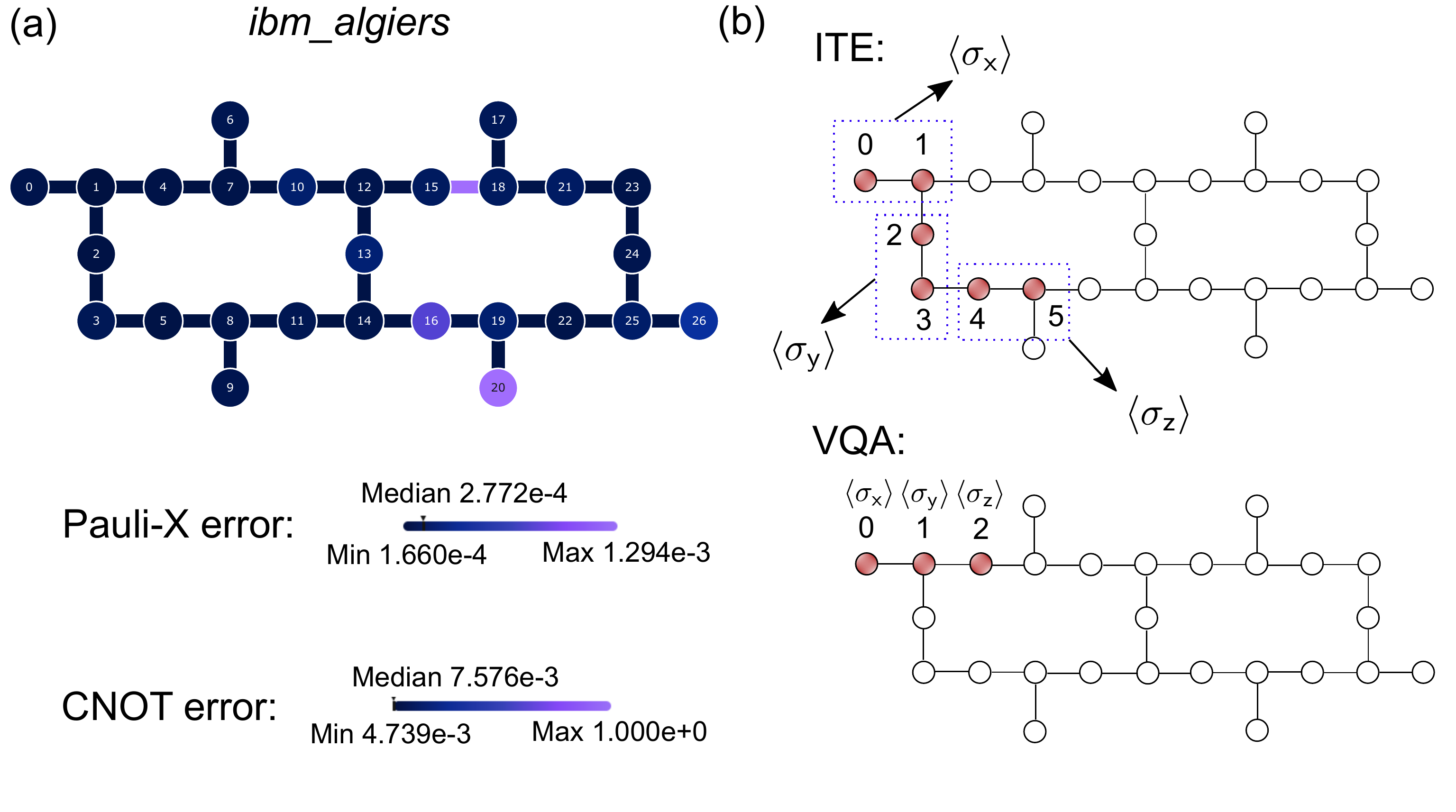}
	\caption{Details of IBM quantum processor: (a) (upper panel) the geometric layout of \texttt{ibm\_algiers} device of the $27$-qubit Falcon type on IBM Q~\cite{IBMprocessortype}. The circles correspond to each qubit, which is connected by a bond. The color indicates the amplitude of different types of gate error (lower pannel): single-qubit Pauli-X error, and two-qubit CNOT gate error. (b) Qubits selection for both ITE and VQA algorithms on \texttt{ibm\_algiers}. For ITE, a total number of $6$ qubits were chosen where qubits $(0,1)$ are for the measurement of $\langle \sigma_x \rangle$, qubits $(2,3)$ for $\langle \sigma_y \rangle$, and qubits $(4,5)$ for $\langle \sigma_z \rangle$, which is consistent with the quantum circuit shown in Fig.~\ref{fig:ITEcircuit}. For VQA, a total number of $3$ qubits were chosen where the first qubit ($0$) is for the measurement of $\langle \sigma_x \rangle$, the second qubit ($1$) for $\langle \sigma_y \rangle$, and the third qubit ($2$) for $\langle \sigma_z \rangle$.}
	\label{fig:aligers qubits}
\end{figure*}

In this section, we outline the details of quantum algorithms as discussed in the main text for using digital quantum computers, i.e., the IBM Q quantum processor.

\subsection{Quantum circuit implementation of two algorithms on IBM Q}

For both methods introduced in the main text, the corresponding circuits are executed on the real IBM Q device on the cloud. Throughout this work, we used a 27-qubit `Falcon'~\cite{IBMprocessortype} processor \texttt{ibm\_algiers} from IBM Q [Fig.~\ref{fig:aligers qubits}(a)]. It is a two-dimensional lattice consisting of single transmon qubits (circles) connected via bonds [see Fig.~\ref{fig:aligers qubits}]. The color on each single qubit and the bond indicates the single-qubit Pauli gate error as well as the two-qubit CNOT gate error, respectively. All circuit submissions are conducted via Qiskit SDK~\cite{Qiskit}.

For the quantum imaginary time evolution (ITE) approach, for each parameter pair $(k_x,k_y)$, it requires $6$ qubits in total, and it requires $4$ quantum circuits to compute all four required projectors $(P_g(k_x,k_y), P_e(k_x,k_y), P_g((k_x+\delta,k_y), P_g(k_x,k_y+\delta)) )$. As the maximum number of circuits to submit to \texttt{ibm\_algiers} is $300$, in order to fully utilize this capacity, a total of $4 \times 60=240$ circuits are submitted for the first epoch of the $60$ parameter pairs of $(k_x,k_y)$ out of the total $15*15=225$ pairs. Then two more epochs are submitted followed by the remaining $4*45=180$ circuits. In this work, we have tested for other different partitions of the circuits and this is the most feasible one to be executed on IBM Q within the walltime provided (less than three hours). The quantum circuit gates decomposition for the unitary operator $\tilde{U}$ for $k_x,k_y=0$ is shown in Fig.~\ref{fig:unitary operator circuit}(b) where there are at most three CNOT gates. Note that in principle, for the imaginary time evolution to obtain the final ground state, the evolution time $\tau$ goes to $\infty$. For all parameters considered in this work, we found that when $\tau=8$, the final state already falls into the true ground state, and therefore we have used $\tau=8$ throughout this work.

For the variational quantum algorithm approach, the pattern of submitting tasks to IBM Q on the could is similar to the ITE approach, with only $3$ qubits utilized. The unitary operator $U$ decomposition from the PQC is plotted in Fig.~\ref{fig:unitary operator circuit}(a) for $k_x,k_y=0$.

Finally, we also remark that for all circuits illustrated in this work, We follow the definitions of $3D$ rotation $U_3$ gates from Qiskit~\cite{Qiskit}:
\begin{align}
\begin{aligned}
\begin{split}U_3(\theta, \phi, \lambda) =
    \begin{bmatrix}
        \cos\left(\frac{\theta}{2}\right)          & -e^{i\lambda}\sin\left(\frac{\theta}{2}\right) \\
        e^{i\phi}\sin\left(\frac{\theta}{2}\right) & e^{i(\phi+\lambda)}\cos\left(\frac{\theta}{2}\right)
    \end{bmatrix}
    \end{split}
    \end{aligned}
\end{align}
where $\theta,\phi,\lambda \in [0,2\pi]$.

\subsection{Measurement and error mitigation with Qiskit Runtime}

The default measurement schemes on IBM Q is in the Pauli-$z$ basis, i.e., measuring $\langle \sigma_z \rangle$, and on IBM Q, the measured outcomes are entirely represented in binary bit strings, i.e. $0$ for spin-up ($|\uparrow\rangle$), and $1$ for spin-down ($|\downarrow \rangle$). Therefore, to calculate $\langle \sigma_z \rangle$, we express it in terms of the normalized counts (or pseudo probability) of spin-up's and spin-down's:

\begin{align}
    \label{eq:magnetizationcalculationonibmq}
    &\langle \psi| \sigma_z |\psi \rangle = \langle \psi |\sigma_{\uparrow} | \psi \rangle -\langle \psi | \sigma_{\downarrow} | \psi\rangle 
\end{align}

where 
$\sigma_{\uparrow}=\begin{bmatrix}
    1 & 0 \\ 0 & 0
\end{bmatrix}$, and $\sigma_{\downarrow}=\begin{bmatrix}
    0 & 0 \\ 0 & 1
\end{bmatrix}$.

To measure other Pauli operator expectation values such as $\langle \sigma_x \rangle$ and $\langle \sigma_y \rangle$, a rotation operator before the measurement is needed:

\begin{align}
    &\langle \psi |\sigma_x | \psi \rangle = \langle \psi | R_y\left(\pi/2\right)\sigma_z R_y\left(-\pi/2\right) | \psi \rangle
\end{align}
and 
\begin{align}
    &\langle \psi |\sigma_y | \psi \rangle = \langle \psi | R_x\left(-\pi/2\right)\sigma_z R_x\left(\pi/2\right) | \psi \rangle
\end{align}
where 
\begin{align}
&R_y(\theta)=\begin{bmatrix}
    \cos{\theta/2} & -\sin{\theta/2}\\
    \sin{\theta/2} & \cos{\theta/2}
\end{bmatrix}, \\ \nonumber
&R_x(\theta)=\begin{bmatrix}
    \cos{\theta/2} & -i \sin{\theta/2}\\
    -i \sin{\theta/2} & \cos{\theta/2}
\end{bmatrix}.
\end{align}
Also, the error mitigation of the readout error (measurement error) has already incorporated into the Qiskit Runtime option~\cite{qiskitruntimeerrormitigation} by setting up the \texttt{resilience\_level} option during the submission of the tasks. Throughout this work, we set the \texttt{resilience\_level} to be $1$, and we found there is not much difference for the major results when setting it to other levels.

Finally, we remark that for all results in this work, the total number of shots we applied is $100000$, i.e., the maximum number of shots which the IBM Q device  \texttt{ibm\_algiers} could offer.

\end{document}